\documentclass[12pt,preprint]{aastex}

\def\gsim{\;\lower4pt\hbox{${\buildrel\displaystyle >\over\sim}$}\;}
\def\lsim{\;\lower4pt\hbox{${\buildrel\displaystyle <\over\sim}$}\;}
\def\grls{\;\lower4pt\hbox{${\buildrel\displaystyle >\over <}$}\;}

\shortauthors{WANG \& ZHANG} \shorttitle{Non-CME Associated Flares}

\begin{document}

\title{A Comparative Study Between Eruptive X-class Flares Associated
with Coronal Mass Ejections and Confined X-class Flares}

\author{Yuming Wang\altaffilmark{1, 2}, and Jie Zhang\altaffilmark{1}}
\altaffiltext{1}{Department of Computational and Data Sciences,
College of Science, George Mason University, 4400 University Dr.,
MSN 6A2, Fairfax, VA 22030, USA, Email: ywangf@gmu.edu,
jzhang7@gmu.edu}

\altaffiltext{2}{School of Earth \& Space Sci., Univ. of Sci. \& Tech. of
China, Hefei, Anhui 230026, China, Email: ymwang@ustc.edu.cn}

\begin{abstract}
Following the traditional naming of ``eruptive flare" and
``confined flares" but not implying a causal relationship between
flare and coronal mass ejection (CME), we refer to the two kinds
of large energetic phenomena occurring in the solar atmosphere as
eruptive event and confined event, respectively: the former type
refers to flares with associated CMEs, while the later type refers
to flares without associated CMEs. We find that about 90\% of
X-class flares, the highest class in flare intensity size, are
eruptive, but the rest 10\% confined. To probe the question why
the largest energy release in the solar corona could be either
eruptive or confined, we have made a comparative study by
carefully investigating 4 X-class events in each of the two types
with a focus on the differences in their magnetic properties. Both
sets of events are selected to have very similar intensity (X1.0
to X3.6) and duration (rise time less than 13 minutes and decaying
time less than 9 minutes) in soft X-ray observations, in order to
reduce the bias of flare size on CME occurrence. We find no
difference in the total magnetic flux of the photospheric source
regions for the two sets of events. However, we find that the
occurrence of eruption (or confinement) is sensitive to the
displacement of the location of the energy release, which is
defined as the distance between the flare site and the
flux-weighted magnetic center of the source active region. The
displacement is 6 to 17 Mm for confined events, but is as large as
22 to 37 Mm for eruptive events, compared to the typical size of
about 70 Mm for active regions studied. In other words, confined
events occur closer to the magnetic center while the eruptive
events tend to occur closer to the edge of active regions.
Further, we have used potential-field source-surface model (PFSS)
to infer the 3-D coronal magnetic field above source active
regions. For each event, we calculate the coronal flux ratio of
low corona ($<$ 1.1 $R_\odot$) to high corona ($\geq$ 1.1
$R_\odot$). We find that the confined events have a lower coronal
flux ratio ($< 5.7$), while the eruptive events have a higher flux
ratio($> 7.1$). These results imply that a stronger overlying
arcade field may prevent energy release in the low corona from
being eruptive, resulting in flares without CMEs. A CME is more
likely to occur if the overlying arcade field is weaker.
\end{abstract}

\section{Introduction}
Coronal mass ejections (CMEs) and flares are known to be the two
most energetic phenomena that occur in the atmosphere of the Sun,
and have profound effects on the physical environment in the
geo-space environment and human technological systems. In this
paper, we intend to understand the physical origins of CMEs and
flares by comparatively studying two different kinds of energetic
phenomena, both of which have almost identical flares, but one is
associated with CMEs and the other not. In the past when lacking
direct CME observations, these two kinds of phenomena have been
called as eruptive flares and confined flares,
respectively~\citep[e.g.,][]{Svestka_Cliver_1992}. An eruptive
flare (also called a dynamic flare) is usually manifested as
having two-ribbons and post-flare loops in H$\alpha$ imaging
observations and long duration (e.g., tens of minutes or hours) in
soft X-ray, while a confined flare occurs in a compact region and
last for only a short period (e.g., minutes). Following this
convention but trying not to imply a causal relation between flare
and CME, hereafter we refer the flare event associated with an
observed CME as an ``eruptive event", and the flare event not
associated with a CME as a ``confined event".

It has been suggested based on observations that CMEs and flares
are the two different phenomena of the same energy release process
in the corona~\citep[e.g.,][]{Harrison_1995, Zhang_etal_2001a,
Harrison_2003}. They do not drive one another but are closely
related. In particular,~\citet{Zhang_etal_2001a}
and~\citet{Zhang_etal_2004} showed that the fast acceleration of
CMEs in the inner corona coincides very well in time with the rise
phase (or energy release phase) of the corresponding soft X-ray
flares, strongly implying that both phenomena are driven by the
same process at the same time, possibly by magnetic field
reconnections. However, this implication raises another important
question of under what circumstances the energy release process in
the corona leads to an eruption, and under what circumstances it
remains to be confined during the process. An answer to this
question shall shed the light on the origin of flares as well as
CMEs.

The occurrence rate of eruptive events depends on the intensity
and duration of flares. A statistical study performed
by~\citet{Kahler_etal_1989} showed that the flares with longer
duration tend to be eruptive, while more impulsive flares tend to
be confined. By using the CME data from Solar Maximum Mission
(SMM) and the flare data from GOES satellites during 1986 --
1987,~\citet{Harrison_1995} found that the association ratio of
flares with CMEs increases from about 7\% to 100\% as the flare
class increases from B-class to X-class, and from about 6\% to
50\% as the duration of flares increases from about 1 to 6
hours.~\citet{Andrews_2003} examined 229 M and X-class X-ray
flares during 1996 -- 1999, and found that the CME-association
rate, or eruptive rate is 55\% for M-class flares and 100\% for
X-class flares. With a much lager sample of 1301 X-ray
flares,~\citet{Yashiro_etal_2005} obtained a similar result that
the eruptive rates of C, M, and X-class flares are $16-25$\%,
$42-55$\%, and $90-92$\%, respectively.~\citet{Yashiro_etal_2005}
work showed that a flare is not necessarily associated with a CME
even if it is as intense as an X-class flare. Such kind of
confined but extremely energetic events were also reported
by~\citet{Feynman_Hundhausen_1994} and~\citet{Green_etal_2002}.

The studies mentioned above indicate the probability of CME
occurrence for a given flare. On the other hand, there is also a
probability of flare occurrence for a given CME. There are CMEs
that may not be necessarily associated with any noticeable X-ray
flares.~\citet{Zhang_etal_2004} reported an extremely gradually
accelerated slow CME without flare association, implying that the
non-flare eruptive event tends to be slowly driven. By Combining
the corona data from SMM and 6-hr soft X-ray data from GOES
satellite,~\citet{St_Cyr_Webb_1991} reported that about 48\%
frontside CMEs were associated with X-ray events near the solar
minimum of solar cycle 21. Based on SOHO
observations,~\citet{Wang_etal_2002a} studied 132 frontside halo
CMEs, and found that the association rate of CMEs with X-ray
flares greater than C-class increased from 55\% at solar minimum
to 80\% near solar maximum. With 197 halo CMEs identified during
1997 -- 2001,~\citet{Zhou_etal_2003} concluded that 88\% CMEs were
associated with EUV brightenings.

Some attempts have been made to explain the confinement or the
eruptiveness of solar energetic events in the context of the
configuration of coronal magnetic field.~\citet{Green_etal_2001}
analyzed the 2000 September 30 confined event utilizing
multi-wavelength data, and suggested that the event involve
magnetic reconnection of two closed loops to form two newly closed
loops without the opening of the involved magnetic
structure.~\citet{Nindos_Andrews_2004} made a statistical study of
the role of magnetic helicity in eruption rate. They found that
the coronal helicity of active regions producing confined events
tends to be smaller than the coronal helicity of those producing
eruptive events.

In this paper, we address the eruptive-confinement issue of solar
energetic events with the approach of a comparative study. We
focus on the most energetic confined events that produce X-class
soft X-ray flares but without CMEs. While the majority of X-class
flares are eruptive, a small percentage (about 10\%) of them are
confined. The magnetic properties of these confined events shall
be more outstanding than those less-energetic confined events. For
making an effective comparison, we select eruptive events with
X-ray properties, in terms of intensity and duration, very similar
to those selected confined events. The differences on magnetic
configuration between these two sets of events shall most likely
reveal the true causes of eruption or confinement. How to select
events and the basic properties of these events are described in
section 2. Detailed comparative analyses of the two sets of events
are given in section~\ref{sec_photosphere} and~\ref{sec_corona},
which are on the properties of the photospheric magnetic field
distribution and the extrapolated coronal magnetic field
distribution, respectively. In section~\ref{sec_summary}, we
summarize the paper.

\section{Selection of Events and Observations}\label{sec_selection}

\subsection{Confined Events: X-class Flares without CMEs}
From 1996 to 2004, there are 104 X-class soft X-ray flares
reported by the NOAA (National Oceanic and Atmosphere
Administration) Space Environment Center (SEC). Flares are
observed by Geosynchronous Earth Observing Satellites (GOES),
which record in high temporal resolution (every 3 seconds) the
disk-integrated soft X-ray flux in two pass-bands: 1.0 -- 8.0~\AA\
and 0.5 -- 4.0~\AA, respectively. The flare catalog provides the
peak intensity, beginning time, peak time, and ending time of
flares. Based on peak intensity, flares are classified into five
categories: A, B, C, M and X in the order of increasing strength.
An X-class flare, in the strongest category, is defined by the
peak flux in the 1.0 -- 8.0~\AA\ band exceeding $10^{-4}$
Wm$^{-2}$.

To find out whether a flare is associated with a CME or not, we
make use of both the CME observations by Large Angle and
Spectrometric Coronagraph (LASCO,~\citealp{Brueckner_etal_1995})
and coronal disk observations by Extreme Ultraviolet Imaging
Telescope (EIT,~\citealp{Delaboudiniere_etal_1995}); both
instruments are on-board the Solar and Heliospheric Observatory
(SOHO) spacecraft. The search process started with the CME
catalog\footnote{the NRL-GSFC-CUA CME catalog at
http://cdaw.gsfc.nasa.gov/CME\_list/}~\citep{Yashiro_etal_2004}
for an initial quick look. A flare becomes a candidate of confined
type if there is no any CME whose extrapolated onset time is
within the 60-minute time window centered at the fare onset time.
The onset time of a CME is calculated by linearly extrapolating
the height-time measurement in the outer corona back to the
surface of the Sun, which shall give the first-order approximation
of the true onset time of CME. Further, we visually examine the
sequence of the LASCO and EIT images around the flare time to
verify that indeed there is no CME associated with the flare
studied. One common property of these confined events is the lack
of EUV dimming in EIT images, even though they show strong compact
brightenings in EIT images. Following the compact brightening,
there is no corresponding CME feature appearing in subsequent
LASCO images. This scenario is in sharp contrast to that of an
eruptive event, in which an EIT dimming accompanies the
brightening, and within a few frames, a distinct CME feature
appears in appropriate position angle in LASCO images. After
applying this process on all X-class flares, we find 11 events are
confined; they are listed in Table~\ref{tb_flares_list}. We notice
that event 7 to 11 occurred within three days between July 15 -
17, 2004, and they all originated from the same solar active region
(NOAA AR10649).

Among the 11 confined X-class flares from 1996 to 2004, the first
four events have been reported earlier
by~\citet{Yashiro_etal_2005}.The third one has also been reported
and studied by~\citet{Green_etal_2002}. During 1996 -- 2004, there
are in total 104 X-class flares. Thus the percentage of confined
X-class flares is about $10\%$. As shown in
Table~\ref{tb_flares_list}, all these confined flares are
impulsive. Their rise time does not exceed 13 minutes except for
event 8 (23 minutes). The decay time does not
exceed 10 minutes for all the 11 events. The rise and decay times
are derived from the begin, maximum and end times of 
flares, which are defined and compiled by 
NOAA/SEC~\footnote{http://www.sec.noaa.gov/ftpdir/indices/events/README}. 
The peak intensity of all
these events was less than X2.0 level except for event 10 (X3.6).
Any event stronger than X3.6 is found always associated with a
CME.

Out of the 11 confined events, we are able to select 4 of them
suitable for further in-depth analysis. These events are numbered
as 4, 5, 6 and 11, respectively. They are suitable because (1) the
flare is isolated, which means that there is no other flare
immediately preceding and following that flare, and (2) there is
no other coronal dimming or CME eruption in the vicinity of the
flare region within a certain period. Events 1 and 3 are not
selected, because they mixed up with a flare-CME pair from the
same source regions. In the presence of an eruptive flare
immediately preceding or following a confined event, we are not
certain how well the confined event is related with the earlier or
later eruptive one. In order to make our analysis as ``clean" as
possible, such events are discarded. Event 2 is also excluded
because its source region is right behind the western limb and
hence no timely magnetogram data is available. Events 7 through 11
were all from the same active region. By over-plotting EIT images
showing flare locations on the MDI magnetogram images, we find
that these flares essentially occurred at the same location within
the active region. Thus we choose only the last event representing
all the five events. The four confined events selected for further
analysis are labelled by $C_1$ through $C_4$ in the second column
of Table~\ref{tb_flares_list}.

\subsection{Eruptive Events: X-class Flares with CMEs}
For making the comparative study with the 4 confined events
mentioned above, a set of four eruptive X-class flares are
selected. These eruptive flares are chosen to have similar
properties in X-rays as those confined events:(1) their rise time
and decay time are less than 13 minutes, (2) their intensities are
between X1.0 to X2.0. Further, their locations are within
$60^\circ$ from solar central meridian in longitude in order to
reduce the projection effect in magnetograms. These flares, which
are relatively impulsive, are indeed associated with CMEs, as
shown in LASCO images. The four events are also listed in
Table~\ref{tb_flares_list} labelled as $E_1$ through $E_4$,
respectively.

\begin{figure}[tb]
  \centering
  \includegraphics[width=0.8\hsize]{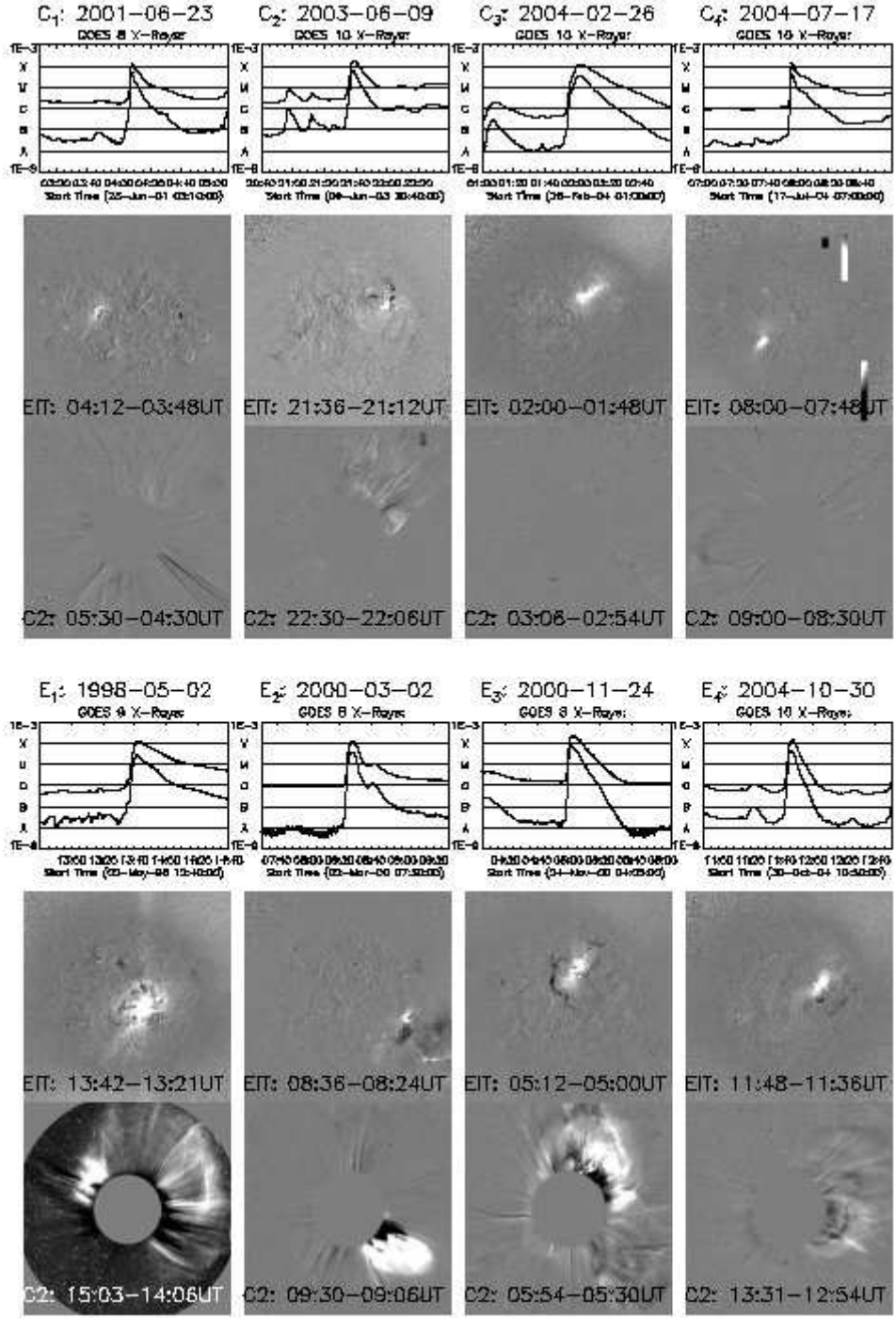}
  \caption{An overview of the four confined ($C_1 - C_4$, upper panels) and four eruptive
($E_1 - E_2$, lower panels) flares. For each events, we display
its GOES X-ray flux profile (spanning 2 hours), running-difference
images of EIT 195\AA\ and LASCO/C2 in the three sub-panels from
top to bottom.}
  \label{fg_flare_overview}
\end{figure}

An overview of the two sets of events is given in
Figure~\ref{fg_flare_overview}. The upper panels show the four
confined events and the lower panels show the four eruptive
events. For each event we show the GOES soft X-ray flux profile
(all in a 2-hour interval), running-difference EIT image and
running-difference LASCO image in the top, middle and low panels,
respectively. Apparently, the temporal profiles of GOES soft X-ray
fluxes exhibit no noticeable difference between the two sets of
events, due to the constraint in our selection of events.
Moreover, as seen in the EIT images, the two sets of events are
all associated with compact coronal brightening indicating the
occurrence of flares. However, for eruptive flares, the
accompanying CMEs are clearly seen in those LASCO images. In
contrast, there is no apparent brightening CME feature seen in
LASCO images for those confined events (only one image is shown
here to represent the observed sequence of images, which all
indicate a non-disturbed corona). For eruptive events, the speeds
and angular widths of CMEs are also listed in the
Table~\ref{tb_flares_list}.

\section{Magnetic Properties in the Photosphere}\label{sec_photosphere}
\subsection{Flare Location and Active Region Morphology}

To explore what physical factors lead the two similar sets of
flares, all strong and impulsive, to have difference in CME
production, we first study the magnetic properties of their
surface source regions. SOHO/MDI provides the observations of
photospheric magnetic field (the component along the line of
sight) every 96 minutes. The spatial resolution of MDI
magnetograms is about 4 arcsec with a plate scale of 2 arcsec per
pixel, at which detailed magnetic features across the source
active regions are reasonably resolved. To reduce the projection
effect of line-of-sight magnetic field, we have only chosen the
events within $60^\circ$ from solar central meridian.

\begin{figure}[tb]
  \centering
  \includegraphics[width=\hsize]{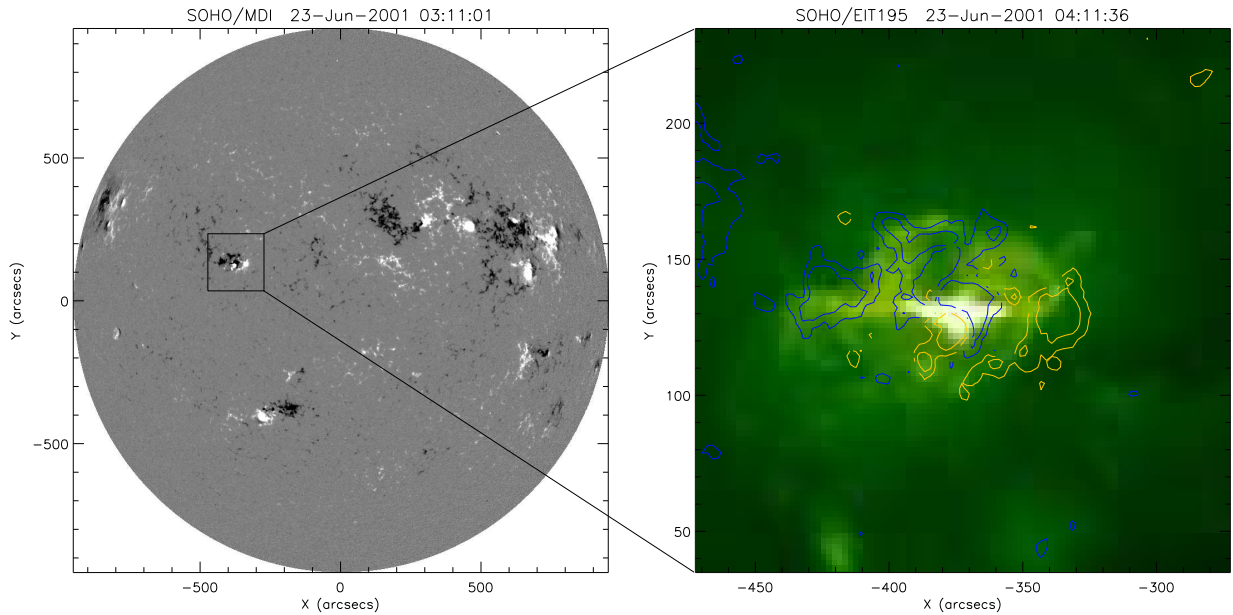}
  \caption{An example showing flare and its source region. The left
image is a full disk MDI magnetogram taken before the flare onset.
The right image shows the EIT image in green-white false colors;
the white patch at the center denotes the flare location. The
superimposed contours show the magnetogram, with yellow the
positive and blue the negative field.}
  \label{fg_pmdieit_example}
\end{figure}

For each event, we determine the location of the flare seen in EIT
relative to the magnetic features seen in MDI. We first align the
MDI image with the corresponding EIT image. The difference in the
timing between MDI and EIT images have been taken into account.
Figure~\ref{fg_pmdieit_example} illustrates the alignment for the
2001 June 23 event. The soft X-ray flare began at 04:02 UT and
peaked at 04:09 UT. In EIT 195~\AA\ image showing the flare was
taken at 04:11 UT. The nearest full disk MDI image prior to the
flare was taken at 03:11 UT. The MDI magnetogram is rotated to fit
the EIT time, and then superimposed in contours onto the EIT
image. In the right panel of Figure~\ref{fg_pmdieit_example}, we
display the aligned images; only the region of interest is shown.
Using this method, we are able to determine the location of
flares, which is just above the neutral lines seen in the
magnetogram.

In Figure~\ref{fg_segmdi_noncme} and~\ref{fg_segmdi_cme}, we show
the magnetogram images for the four confined events and the four
eruptive events, respectively. The flare sites, or bright patches
seen in EIT, are marked by red asterisks in the images. The
magnetogram images have been re-mapped onto the Carrington
coordinate, which reduces the spherical projection effect of the
image area. The $x$-axis is Carrington longitude in units of
degree, and the $y$-axis is the sine of latitude. The area of the
images shown in Figure~\ref{fg_segmdi_noncme}
and~\ref{fg_segmdi_cme} are all $30^\circ\times30^\circ$ squares,
which usually cover the entire active regions producing the
CMEs/flares of interest. To highlight the magnetic features, the
displayed images have been segmented into three different levels:
strong positive magnetic field ($\geq 50$ Gauss) indicated by
white color, strong negative magnetic field ($\leq -50$ Gauss)
indicated by black color, and weak field (from $-50$ to 50 Gauss)
indicated by gray color. Note that the noise level of a MDI
magnetogram image is typically at about 10 Gauss. As shown in the
figures, an active region is naturally segmented into many
individual pieces. Those pieces with magnetic flux larger than
$10^{13}$ wb are labelled by a letter with a number in the
neighbouring bracket indicating the magnetic flux in units of
$10^{13}$ wb.

\subsection{Results}
We find that there is no apparent difference in term of total
magnetic flux of the source region between the confined events and
eruptive events. The total magnetic flux, combining both positive
and negative flux, are listed in Table~\ref{tb_photosphere}. The
total flux for confined events varies from about 5 to
36$\times10^{13}$ wb, while for eruptive events it varies from
about 11 to 24$\times10^{13}$ wb.

\begin{figure}[tb]
  \centering
  \includegraphics[width=\hsize]{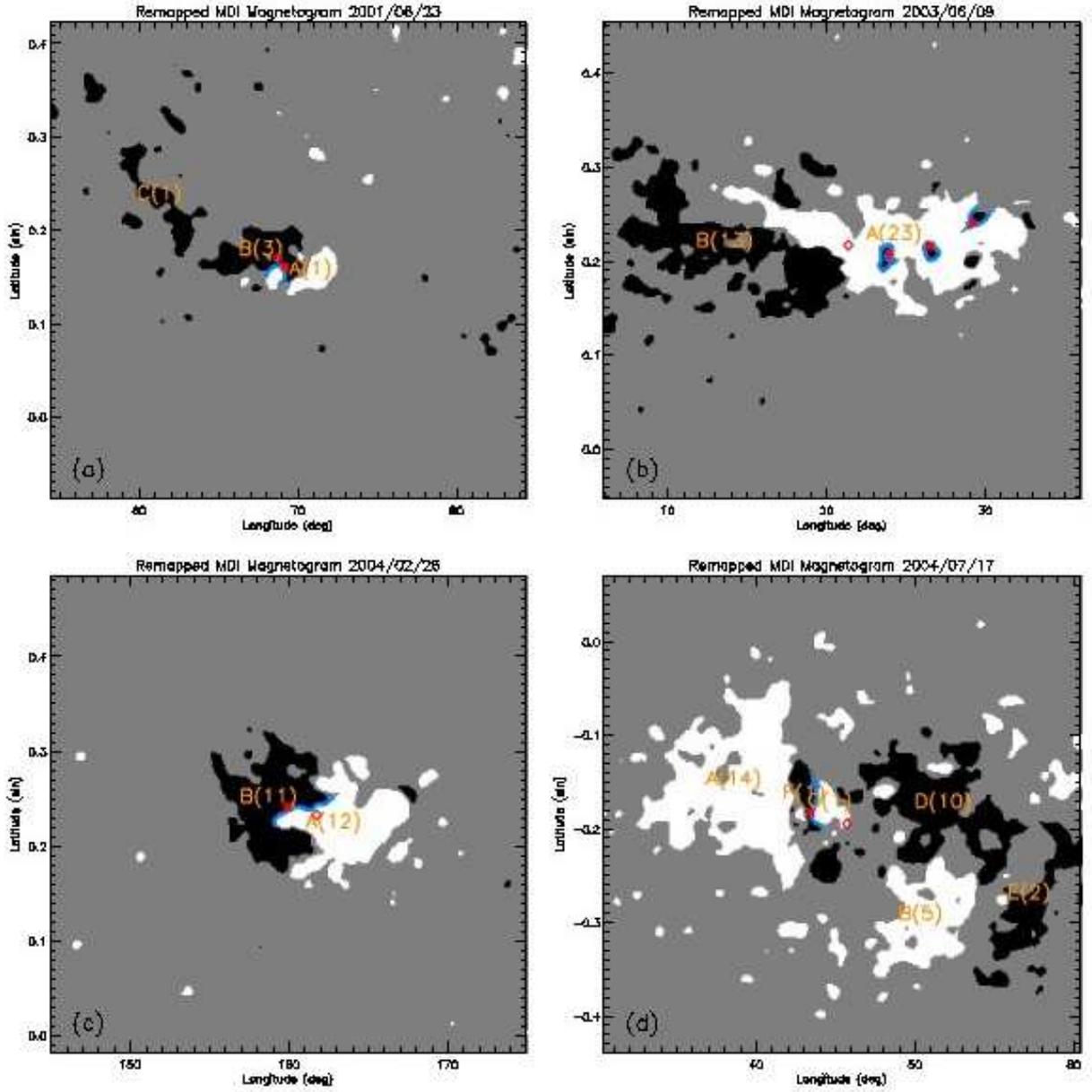}
  \caption{Segmented MDI magnetograms for the four confined events, in which
strong positive ($\geq 50$ Gauss) and negative ($\leq -50$ Gauss)
magnetic fields are highlighted as white and black colors,
respectively. Red asterisk symbols indicate the flare sites, the
red diamond symbols indicate the COM of the active regions, and
the blue lines denote the neutral lines over which the flares
occurred. See text for more details.}
  \label{fg_segmdi_noncme}
\end{figure}

However, there is a noticeable pattern that the confined flares
all originated in a location relatively closer to the center of
the host active region. Figure~\ref{fg_segmdi_noncme}a shows the
confined event of 2001 June 23. There are three relatively large
magnetic pieces labelled by `A', `B' and `C'. The flare site is
surrounded by the pieces `A' and `B'.
Figure~\ref{fg_segmdi_noncme}b shows the 2003 August 9 confined
event. The flare location is associated with three small negative
patches (marked by the red asterisks) embedded in a large positive
piece `A'. Figure~\ref{fg_segmdi_noncme}c shows the 2004 February
26 confined event. The flare occurred just above the neutral line
between thye large positive piece `A' and the large negative piece
`B'. Figure~\ref{fg_segmdi_noncme}d shows the 2004 July 17
confined event. The flare was located in a complex active region
with a large number of sunspots. It occurred right at the boundary
between pieces `C' and `F'. From the view of the entire active
region, pieces `C' and `F' were further enclosed by two much
larger and stronger pieces `A' and `D' whose fluxes were about 10
times larger.

\begin{figure}[tb]
  \centering
  \includegraphics[width=\hsize]{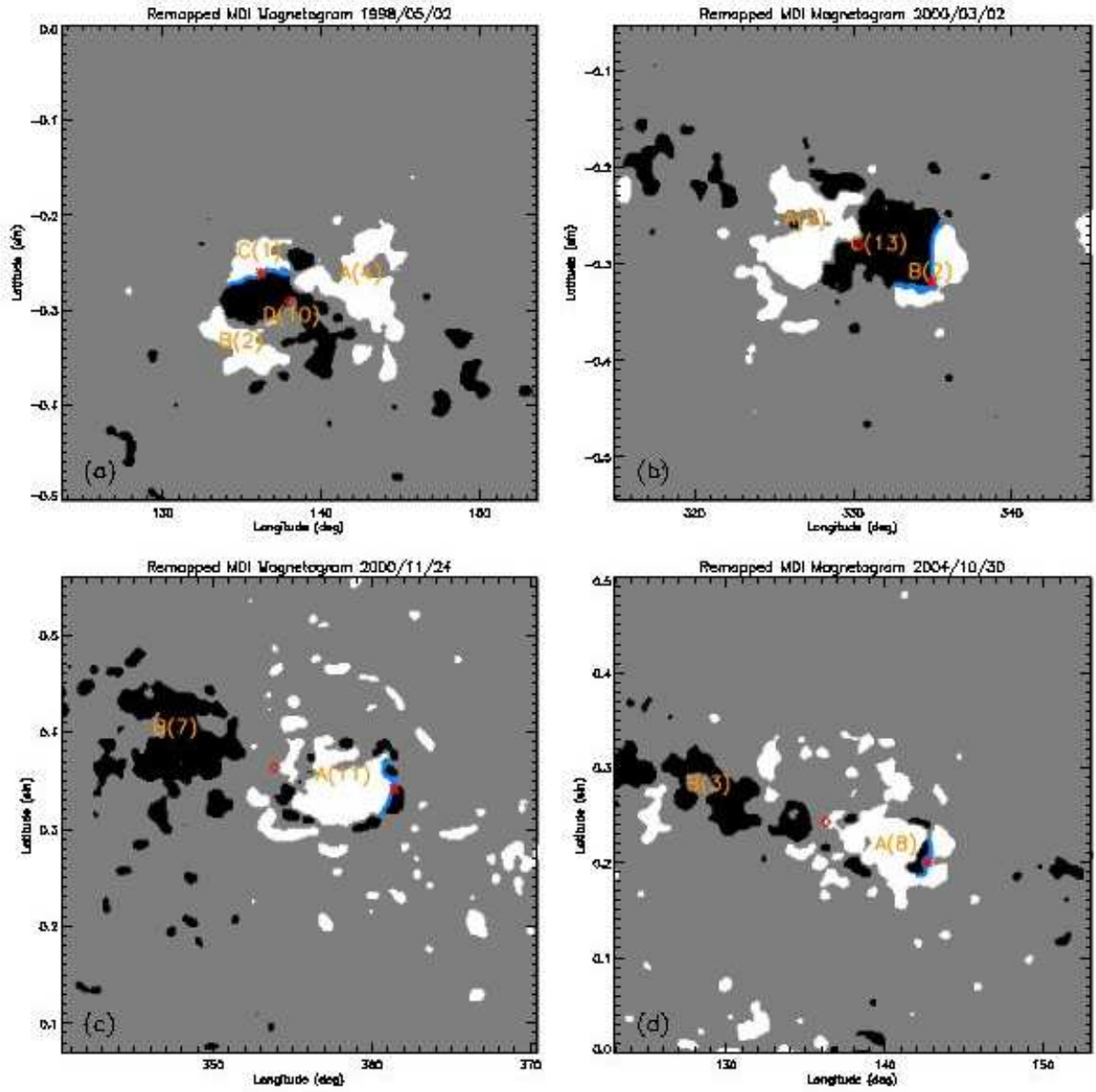}
  \caption{Segmented MDI magnetograms for the four eruptive events. See the caption in Figure~\ref{fg_segmdi_noncme}.}
  \label{fg_segmdi_cme}
\end{figure}

For those eruptive flares, on the other hand, the flare sites were
all relatively further from the center of the magnetic flux
distribution. In other words, they were closer to the edge of
hosting active regions. Figure~\ref{fg_segmdi_cme}a shows the 1998
May 2 event. The strongest pair of magnetic pieces are `A' and
`D', but the eruptive flare occurred at the neutral line between
pieces `D' and `C', which was the smallest amongst the 4 labelled
pieces in the active region. Figure~\ref{fg_segmdi_cme}b shows the
2000 March 2 event. Similarly, the strongest pair of pieces were
`A' and `C', but the flare was from the neutral line between
pieces `C' and `B', which was the smallest labelled piece.
Figure~\ref{fg_segmdi_cme}c shows the 2000 November 24 event. The
flare occurred at the outer edge of the strongest piece `A', which
was neighbored by a very small region with negative flux.
Figure~\ref{fg_segmdi_cme}d shows the 2004 October 30 event. The
flare site was also close to the edge of the entire active region.

To quantify this observation of different displacements of flare
locations, we here introduce a flare displacement parameter, which
is defined by the surface distance between the flare site and the
weighted center of the magnetic flux distribution of the host
active region, or center of magnetic flux (COM) for short. The COM
might be the place that has the most overlying magnetic flux. The
COM is calculated based on the re-mapped $30^\circ\times30^\circ$
MDI images (without segmentation). It is a point, across which any
line can split the magnetogram into two flux-balanced halves, and
can be formulated as $x_c=\frac{\sum_iF_i*x_i}{\sum_iF_i}$ and
$y_c=\frac{\sum_iF_i*y_i}{\sum_iF_i}$. The COM of these events
have been marked by the diamonds in Figure~\ref{fg_segmdi_noncme}
and~\ref{fg_segmdi_cme}. With known COM, it is easy to derive the
displacement parameter, which is listed in
Table~\ref{tb_photosphere}. Consistent with the earlier
discussions, it is found that the displacement parameters for the
four confined events are all smaller than 17 Mm, while for the
four eruptive events, they are all larger than 22 Mm.

\begin{figure}[tb]
  \centering
  \includegraphics[width=\hsize]{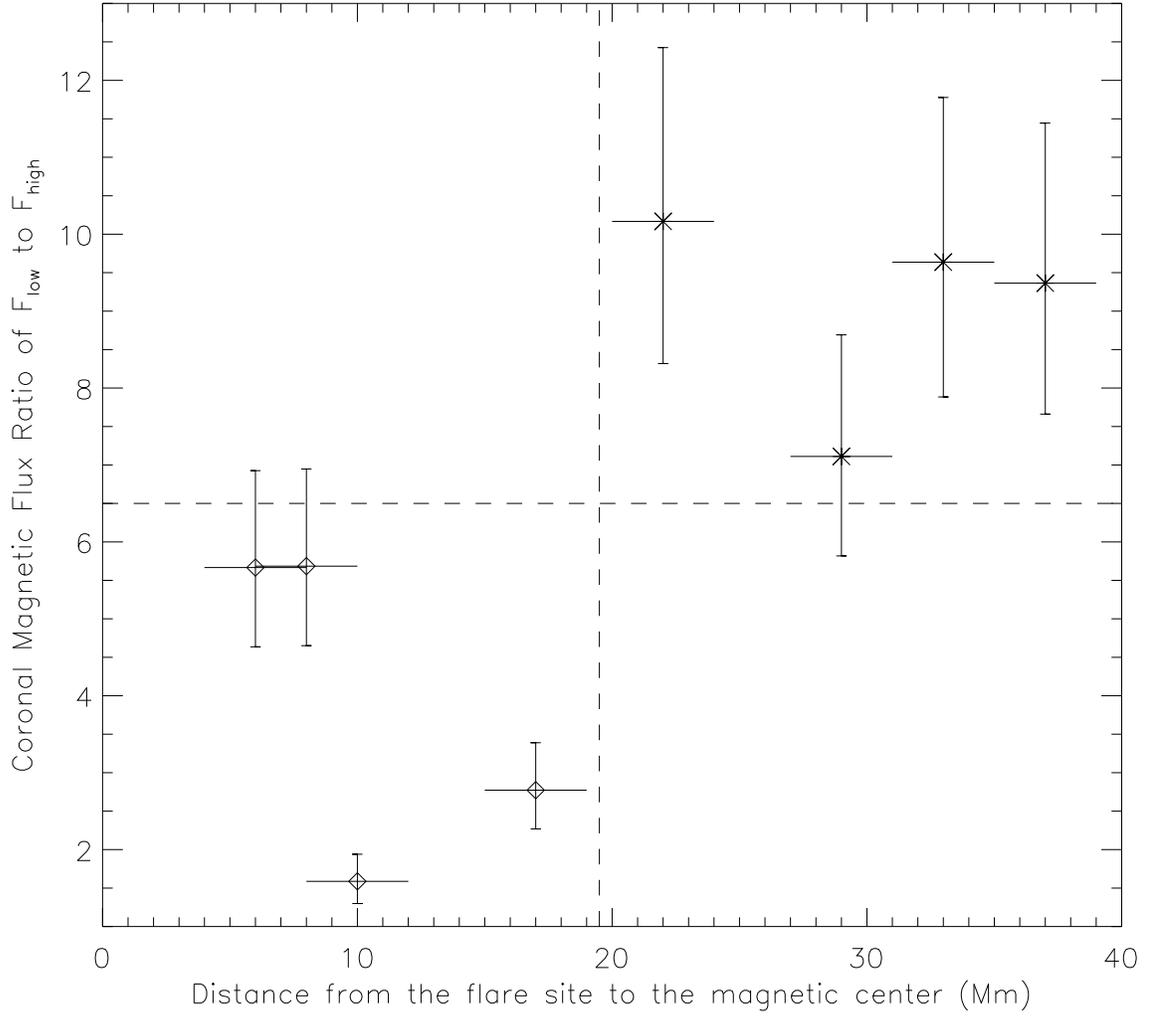}
  \caption{The scattering plot showing magnetic properties of
both confined events (diamond symbols) and the eruptive events
(asterisk symbols). The $x$-axis denotes the distance between the
flare site and the center of magnetic flux distribution (COM) of
the active region, and the $y$-axis denotes the ratio of magnetic
flux in the low to high corona above the neutral line.}
  \label{fg_ratio_distance}
\end{figure}

We now consider possible errors in calculating the displacement
parameter. The error main arises from the uncertainty in the
recorded weak magnetic field around the active regions. However,
in the selected region of study that is $30^\circ\times30^\circ$
across, the highlighted white and black pieces contain about 99\%
of the total magnetic flux in the region. Therefore the
uncertainty of the flux is expected to be at the order of 1\%.
Considering the formula of the coordinates of COM given in the
last paragraph and assuming a typical scale of 100 Mm, the error
of the calculated distance is about 1 Mm. Further, considering the
spatial resolution of MDI of $\sim1$ Mm (varing from $\sim0.7$ Mm
at central meridian to $\sim1.4$ Mm at longitude of
$\pm60^\circ$), the overall uncertainty should be about $\pm2$ Mm.
With these consideration, the displacement parameters and their
uncertainties are plotted in Figure~\ref{fg_ratio_distance}. The
confined events are indicated by diamonds, and the eruptive events
are indicated by asterisks. The vertical dashed line, which
corresponds to a displacement of about 19.5 Mm, effectively
separates the two sets of events.

\section{Magnetic Properties in the Corona}\label{sec_corona}
\subsection{Method}
Having studied the magnetic field distributions in the
photosphere, we further investigate into magnetic field
distributions in the 3-D corona. The magnetic field configuration
in the corona shall ultimately determine the eruption/confinement
since the energy releases occurs in the corona. There is so far no
direct observations of coronal magnetic fields. We have to utilize
certain models to calculate the coronal magnetic field based on
observed photospheric boundary. In this paper, we apply the
commonly used potential-field source-surface (PFSS)
model~\citep[e.g.,][]{Schatten_etal_1969, Hoeksema_etal_1982}. The
PFSS model is thought to be a useful first-order approximation to
the global magnetic field of the solar corona. Nevertheless, we
realize that the current-carrying core fields, which are low-lying
and near the magnetic neutral line, are far from the potential
field approximation; these core fields are likely to be the
driving source of any energy release in the corona. Therefore, the
usage of PFSS model in this study is limited to calculate the
total flux of the overlying large scale coronal field, which is
believed to be closer to a potential approximation. These
overlying fields are thought to constrain the low-lying field from
eruption.

A modified MDI magnetic field synoptic chart is used as input to
our PFSS model. The high-resolution MDI magnetic field synoptic
chart\footnote{http://soi.stanford.edu/magnetic/index6.html} is
created by interpolating data to disk-center resolution, resulting
in a $3600\times1080$ pixel map. The X and Y axis are linear in
Carrington longitude ($0.1^\circ$ intervals) and sine latitude,
respectively. This high resolution is useful in creating detailed
coronal magnetic field above the source regions of interest. Since
an MDI synoptic chart is created from the magnetogram images over
a $\sim27$-day solar rotation, the synoptic chart does not exactly
represent the photospheric magnetic field in the region of
interest at the flare/CME time. The details of the source region
may be different because of the evolution of photospheric magnetic
field. To mitigate this problem, we use the MDI daily magnetogram
to update the original MDI synoptic chart. The process is to
re-map the snapshot magnetogram image prior to the flare
occurrence to the Carrington grid, and then slice out the region
of interest, $30^\circ$ in longitude and $60^\circ$ in latitude.
This sliced region, to replace the corresponding portion in the
original synoptic chart.

Since a PFSS model makes use of the spheric harmonic series
expansion, we realize that a high-resolution data requires a high
order expansion in order to have a consistent result. We calculate
the spheric harmonic coefficients to as high as 225 order for the
input $3600\times1080$ boundary image. We find that, at this
order, we can get the best match between the calculated
photospheric magnetic field and the input synoptic chart. The mean
value of the difference between them is less than 0.5 Gauss, and
the standard deviation is generally $\lsim 15$ Gauss for solar
minimum and $\lsim 25$ Gauss for solar maximum, which are
comparable to the noise level of MDI magnetograms. That means, we
can effectively reproduce the observed photospheric magnetic field
with the 225-order PFSS model. With spheric harmonic coefficients
known, it is relatively straightforward to calculate the magnetic
field in the 3-D volume of the corona.

\subsection{Results}
\begin{figure}[tb]
  \centering
  \includegraphics[width=\hsize]{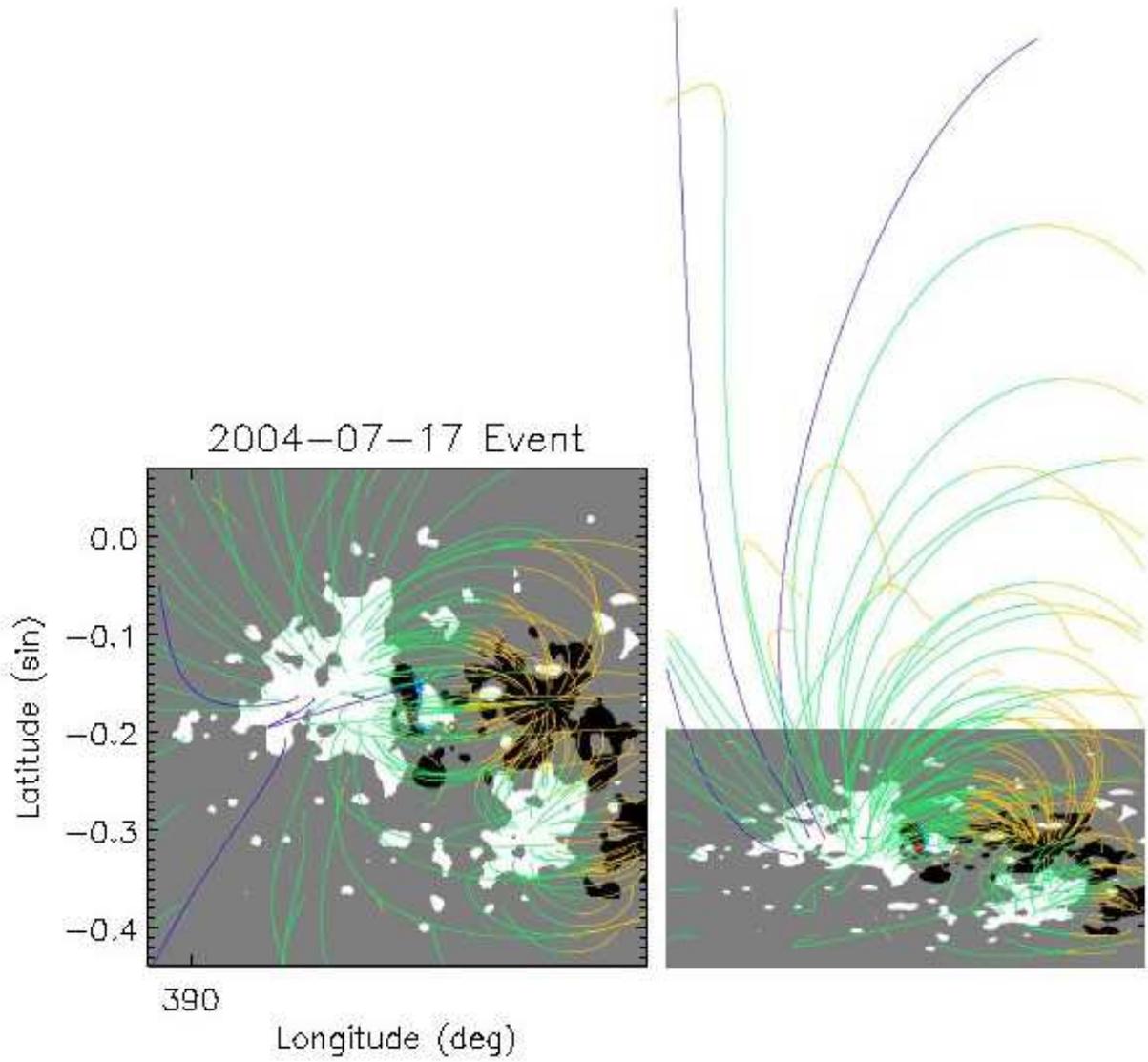}
  \caption{Calculated coronal magnetic field of one confined event.
The closed field lines are denoted by the green \& yellow colors,
corresponding to the outward and inward direction respectively,
and the open field lines are denoted by the blue color. The left
image is of a top view while the right image is of a side view.}
  \label{fg_mag_n_20040717}
\end{figure}

\begin{figure}[tb]
  \centering
  \includegraphics[width=\hsize]{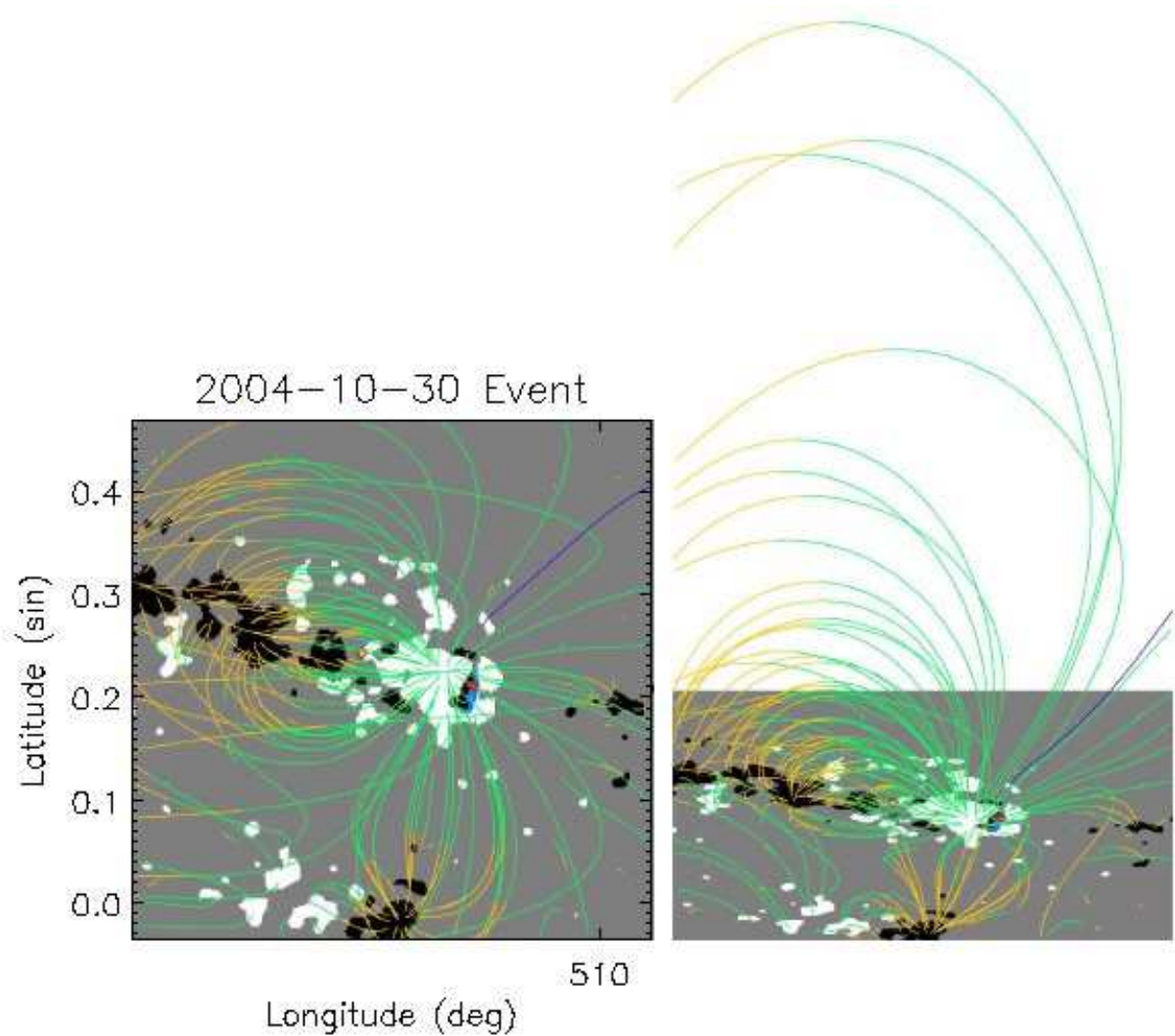}
  \caption{An example of eruptive events showing the extrapolated magnetic
field above the active region.}
  \label{fg_mag_c_20041030}
\end{figure}

Figure~\ref{fg_mag_n_20040717} and~\ref{fg_mag_c_20041030} show
the extrapolated coronal magnetic field lines for one confined
event (2004 July 17) and one eruptive event (2004 October 30),
respectively. In each figure, the left panel shows field lines
viewed from top, and the right panel shows field lines viewed from
the side by rotating the left panel view of $60$ degrees into the
paper. The green-yellow colors denote the closed field lines, with
green indicating the loop part of outward magnetic field (positive
magnetic polarity at the footpoints) and yellow the inward
(negative magnetic polarity at the footpoints), and the blue color
indicates the open field lines. The two examples show that the
location of the confined flare, which is near the center of the
active region, is covered by a large tuft of overlying magnetic
loop arcades, while the location of the eruptive flare, which is
near the edge of the active region, has relatively few directly
overlying loop arcade. In particular, for the eruptive event, the
nearby positive and negative magnetic field lines seem to connect
divergently with other regions, instead of forming a loop arcade
of its own.

To quantify the strength of the overlying field, we calculate the
total magnetic flux cross the plane with the $x$ direction
extending along the neutral line and the $y$ direction vertically
along the radial direction. The thick blue lines on the
photospheric surface in Figure~\ref{fg_segmdi_noncme}
and~\ref{fg_segmdi_cme} indicate the neutral lines used in the
calculation. The length of the neutral lines is determined as it
encompasses the major part of the eruption region. The overlying
magnetic field flux then is normalized to the length of the
neutral line. Thus obtained normalized overlying magnetic flux is
a better quantity to be used for comparison between different
events, because this parameter is not sensitive to the exact
length of the neutral lines selected, which may vary
significantly from event to event.

Such calculated magnetic fluxes for the 8 events are listed in
Table~\ref{tb_flux}. In calculating the flux, we do not consider
that the crossing direction of the field lines over the neutral
line is from one side to the other or opposite. The relative
uncertainty of the calculated magnetic flux can be estimated as
$\sigma_B/B_0$ where $\sigma_B$ is the uncertainty of calculated
magnetic field strength in the corona and $B_0$ is the magnetic
field strength in the active regions at photosphere. Considering
$\sigma_B$ is about 15 to 25 Gauss, the standard deviation
mentioned before, and $B_0$ is usually hundreds of Gauss, we infer
that the uncertainty of overlying magnetic flux is about 10\%.
The estimate should be true in case that the coronal magnetic field
is correctly obtained by our extrapolation method. If the 
extrapolated field largely deviates from the realistic status, the
uncertainty may probably be slightly different.

The total overlying flux, $F_{total}$, in the height range from
1.0 to 1.5 $R_\odot$, is given in the third column. It seems that
there is no systematic difference between the confined events and
eruptive events. The value for the confined events varies from
0.40 to 1.27 $\times 10^{10}$ wb/Mm, and that for the eruptive
events from 0.73 to 1.34 $\times10^{10}$ wb/Mm.

We further calculate the flux in two different height ranges, the
lower flux from 1.0 to 1.1 $R_\odot$ and the higher flux from 1.1
to 1.5 $R_\odot$. A common accepted scenario is that the lower
flux shall correspond to the inner sheared core field (or
fully-fledged flux rope if filament present) that tends to move
out, while the outer flux is the large scale overlying arcade that
tends to constrain the inner flux from eruption. Note that the
chosen of 1.1 $R_\odot$, which corresponds to a height of about 70
Mm above the surface, is rather arbitrary. However, slightly
changing this number will not affect any overall results that will
be reached. The magnetic flux in the low corona, $F_{low}$, and in
the high corona, $F_{high}$, are listed in the fourth and fifth
columns of Table~\ref{tb_flux}, respectively.

There is a trend that the low-corona flux for the eruptive events
is generally larger than that for the confined events. Three out
of the four eruptive events have their low-corona overlying flux
more than 1.0 $\times10^{10}$ wb/Mm, while three out of the four
confined events have the flux less than 1.0 $\times10^{10}$ wb/Mm.
On the other hand, the high-corona flux for the eruptive events
seems smaller than that for the confined events. Three confined
events have their high-corona flux more than 0.15 $\times10^{10}$
wb/Mm, while all four eruptive events have the flux less than 0.15
$\times10^{10}$ wb/Mm.

We further calculate the flux ratio parameter, which is defined as
$R=F_{low}/F_{high}$. This quantity is independent of the normalization.
It may serve as an index of how weak the
constraint on the inner eruptive field is. Interestingly, the flux
ratios for the two sets of events fall into two distinct groups.
For the confined events, $R$ varies from 1.59 to 5.68, while for
the eruptive events, the value of $R$ is larger, from 7.11 to
10.17. The value of 6.5 may be used as a boundary separating the
two sets of events. This value probably implies a threshold for
confinement or eruptiveness. This is to say, if the flux ratio is
less than 6.5, a flare is likely to be confined, otherwise
eruptive. The higher the ratio, the higher the possibility of a
coronal energy release being eruptive.

\section{Summary and Discussions}\label{sec_summary}
In summary, among the 104 X-class flares occurred during 1996 --
2004, we found a total of 11 ($\sim10\%$) are confined flares
without associated CMEs, and all the others ($\sim90\%$) are
eruptive flares associated with CMEs. Four suitable confined
flares are selected to make a comparative study with four eruptive
flares, which are similar in X-ray intensity and duration as those
confined events. We have carefully studied the magnetic properties
of these events both in the photosphere and in the corona. The
following results are obtained:

(1) In the photosphere, we can not find a difference of the total
magnetic fluxes in the surface source regions between the two sets
of events. However, there is an apparent difference in the
displacement parameter, which is defined as the surface distance
between the flare site and the center of magnetic flux
distribution. For the confined events, the displacement is from 6
to 17 Mm, while for those eruptive events it is from 22 to 37 Mm.
This result implies that the energy release occurring in the
center of an active region is more difficult to have a complete
open eruption, resulting in a flare without CME. On the other
hand, the energy release occurring away from the magnetic center
has a higher probability to have an eruption, resulting in both
flares and CMEs. Whether an eruption could occur or not may be
strongly constrained by the overlying large scale coronal magnetic
field. The overlying coronal magnetic field shall be strongest and
also longest along the vertical direction over the center of an
active region. On the other hand, the overlying constraining field
shall be weaker if the source is away from the center. This
scenario is further supported by our study of coronal magnetic
field.

(2) Calculation of coronal magnetic field shows that the flux
ratio of the magnetic flux in the low corona to that in the high
corona is systematically larger for the eruptive events than that
for the confined events. The magnetic flux ratio for the confined
events varies from 1.6 to 5.7, while the ratio for the eruptive
events from 7.1 to 10.2. However, there is no evident difference
between the two sets of events in the total magnetic flux straddling 
over neutral lines, and there is only a weak trend indicating a 
systematic difference in the low- and high-corona magnetic fluxes.
This low-to-high corona magnetic flux ratio serves as a proxy of
the strength of the inner core magnetic field, which may play an
erupting role, relative to the strength of the overlying large
scale coronal magnetic field, which may play a constraining role
to prevent eruption. The lower this ratio, the more difficult the
energy release in the low corona can be eruptive.

There is variety of theoretical models on the initiation mechanism
of CMEs and the energy release of flares
~\citep[e.g.,][]{Sturrock_1989, Chen_1989,
vanBallegooijen_Martens_1989, Forbes_Isenberg_1991,
Moore_Roumeliotis_1992, Low_Smith_1993, Mikic_Linker_1994,
Antiochos_etal_1999, Lin_Forbes_2000}. These models differ in
pre-eruption magnetic configurations, trigger processes, or where
magnetic reconnection occurs. Nevertheless, in almost all these
models, the magnetic configuration involves two magnetic regimes,
one is the core field in the inner corona close to the neutral
line, the other is the large scale overlying field or background
field. The core field is treated as highly sheared or as a
fully-fledged flux rope; in either case, the core field stores
free energy for release. On the other hand, the overlying field is
regarded as potential and considered to be main constraining force
to prevent the underlying core field from eruption or
escaping.~\citet{Torok_Kliem_2005} and~\citet{Kliem_Torok_2006} 
recently pointed out that the
decrease of the overlying field with height is a main factor in
deciding whether the kink-instability (in their twist flux rope
model) leads to a confined event or a CME. On the other 
hand,~\citet{Mandrini_etal_2005} reported the smallest CME event ever 
observed by 2005, in which the CME originated from the smallest source 
region, a tiny dipole, and developed into the smallest magnetic cloud.
They suggested that the ejections of tiny flux ropes are possible.
Therefore, it is reasonable to argue that, whether an energy release in the corona
is eruptive or confined, is sensitive to the balance between the
inner core field and the outer overlying field. Our observational
results seem to be consistent with this scenario.

This study is only a preliminary step to investigate the
confinement and/or eruptiveness of solar flares, or coronal energy
releases in general. However, it demonstrates that the
distribution of magnetic field both in the photosphere and in the
corona may effectively provide the clue of the possible nature of
an energetic event: whether a flare, a CME or both. To further
evaluate the effectiveness of this methodology, a more robust
study involving more events is needed.

\acknowledgments{We acknowledge the use of the solar data from the
LASCO, EIT and MDI instruments on board SOHO spacecraft. The
SOHO/LASCO data used here are produced by a consortium of the
Naval Research Laboratory (USA), Max-Planck-Institut fuer
Aeronomie (Germany), Laboratoire d'Astronomie (France), and the
University of Birmingham (UK). SOHO is a project of international
cooperation between ESA and NASA. We also acknowledge the use of
CME catalog generated and maintained at the CDAW Data Center by
NASA and The Catholic University of America in cooperation with
the Naval Research Laboratory, and the solar event reports
compiled by the Space Environment Center of NOAA. We thank the
useful discussion with X. P. Zhao at Stanford University, who
provides the procedure of PFSS model. This work is supported by
NSF SHINE grant ATM-0454612 and NASA grant NNG05GG19G. Y. Wang is
also supported by the grants from NSF of China (40525014) and 
MSTC (2006CB806304), and J. Zhang is also supported by NASA 
grants NNG04GN36G.}


\clearpage

\begin{table}[t]
\linespread{1.5} \caption{List of Confined X-class Flares from
1996 to 2004 and Selected Eruptive Flares} \label{tb_flares_list}
\tabcolsep 2pt
\footnotesize 
\begin{tabular}{lcccccccccp{100pt}}
\hline
\# &Label &Date &Begin &$T_R^a$ &$T_D^b$ &Class &Location &NOAA &CME$^c$ &Comment \\
& & &UT &min &min & & &AR &V(km/s)/Width & \\
\hline
\multicolumn{11}{c}{Confined Flares}\\
1     & &2000/06/06 &13:30 &9.0  &7.0  &X1.1 &N20E18 &9026  &- &Contained by a preceding and a following M-class flares (Y)\\
2     & &2000/09/30 &23:13 &8.0  &7.0  &X1.2 &N07W91 &9169  &- &Limb event (G, Y)\\
3     & &2001/04/02 &10:04 &10.0 &6.0  &X1.4 &N17W60 &9393  &- &Contained by a preceding eruptive flare (Y)\\
4 &$C_1$ &2001/06/23 &04:02 &6.0  &3.0  &X1.2 &N10E23 &9511  &- &(Y) \\
5 &$C_2$ &2003/06/09 &21:31 &8.0  &4.0  &X1.7 &N12W33 &10374 &- & \\
6 &$C_3$ &2004/02/26 &01:50 &13.0 &7.0  &X1.1 &N14W14 &10564 &- & \\
7     & &2004/07/15 &18:15 &9.0  &4.0  &X1.6 &S11E45 &10649 &- & \\
8     & &2004/07/16 &01:43 &23.0 &6.0  &X1.3 &S11E41 &10649 &- & \\
9     & &2004/07/16 &10:32 &9.0  &5.0  &X1.1 &S10E36 &10649 &- & \\
10    & &2004/07/16 &13:49 &6.0  &6.0  &X3.6 &S10E35 &10649 &- & \\
11&$C_4$ &2004/07/17 &07:51 &6.0  &2.0  &X1.0 &S11E24 &10649 &- &Event 7--11 all from the same AR \\
\hline
\multicolumn{11}{c}{Eruptive Flares}\\
1 &$E_1$ &1998/05/02 &13:31 &11.0 &9.0  &X1.1 &S15W15 &8210  &936/halo       &\\
2 &$E_2$ &2000/03/02 &08:20 &8.0  &3.0  &X1.1 &S18W54 &8882  &776/62$^\circ$ &\\
3 &$E_3$ &2000/11/24 &04:55 &7.0  &6.0  &X2.0 &N19W05 &9236  &1289/halo      &\\
4 &$E_4$ &2004/10/30 &11:38 &8.0  &4.0  &X1.2 &N13W25 &10691 &427/halo       &\\
\hline
\end{tabular}\\
$^a$ Rise time of flares.\\
$^b$ Decay time of flares.\\
$^c$ Apparent speed and angular width of CMEs. Adopted from the online GSFC-NRL-CUA CME catalog.\\
G and Y in comment column mean that the corresponding events have
been reported by~\citet{Green_etal_2002}
and~\citet{Yashiro_etal_2005}, respectively.
\end{table}

\clearpage

\begin{table}[t]
\linespread{1.5} \caption{Magnetic Properties of the Source Active
Regions of the Confined and Eruptive Flares}
\label{tb_photosphere} \tabcolsep 2pt
\footnotesize 
\begin{tabular}{lccc}
\hline
Event &Date &Flux$^a$ &Distance$^b$ \\
& &$10^{13}$ wb &Mm \\
\hline
\multicolumn{4}{c}{Confined Flares} \\
$C_1$ &2001/06/23 &5 &6 \\
$C_2$ &2003/06/09 &36 &17 \\
$C_3$ &2004/02/26 &23 &8 \\
$C_4$&2004/07/17 &34 &10 \\
\multicolumn{4}{c}{Eruptive Flares} \\
$E_1$ &1998/05/02 &17 &22 \\
$E_2$ &2000/03/02 &24 &33 \\
$E_3$ &2000/11/24 &18 &37 \\
$E_4$ &2004/10/30 &11 &29 \\
\hline
\end{tabular}\\
$^a$ Total magnetic flux in active regions measured in MDI magnetogram.\\
$^b$ Surface distance between the flare site and the COM of the
associated active region.
\end{table}

\clearpage

\begin{table}[t]
\linespread{1.5} \caption{Magnetic flux per unit length overlying the neutral lines}
\label{tb_flux} \tabcolsep 2pt
\footnotesize 
\begin{tabular}{lccccc}
\hline
Event &Date &$F_{total}$ &$F_{low}$ &$F_{high}$ &Ratio$=\frac{F_{low}}{F_{high}}$ \\
& &$10^{10}$ wb/Mm &$10^{10}$ wb/Mm &$10^{10}$ wb/Mm & \\
\hline
\multicolumn{6}{c}{Confined Flares} \\
$C_1$ &2001/06/23 &0.40 &0.34 &0.06 &5.67 \\
$C_2$ &2003/06/09 &0.83 &0.61 &0.22 &2.77 \\
$C_3$ &2004/02/26 &1.27 &1.08 &0.19 &5.68 \\
$C_4$ &2004/07/17 &1.19 &0.73 &0.46 &1.59 \\
\multicolumn{6}{c}{Eruptive Flares} \\
$E_1$ &1998/05/02 &1.34 &1.22 &0.12 &10.17 \\
$E_2$ &2000/03/02 &1.17 &1.06 &0.11 & 9.64 \\
$E_3$ &2000/11/24 &1.14 &1.03 &0.11 & 9.36 \\
$E_4$ &2004/10/30 &0.73 &0.64 &0.09 & 7.11 \\
\hline
\end{tabular}
\end{table}

\clearpage

\clearpage

\clearpage

\clearpage

\clearpage

\clearpage

\clearpage

\end{document}